\documentclass[10pt,journal,twoside]{IEEEtran}
\usepackage{multirow}
\usepackage{amssymb}
\usepackage[dvips]{graphicx}
\usepackage{diagbox}
\usepackage{amsmath}
\usepackage{amsthm}
\usepackage{latexsym,bm}
\usepackage{color}
\usepackage{subfigure}
\usepackage{longtable}
\usepackage{accents}
\usepackage{cite}
\usepackage{enumerate}
\usepackage{arydshln}
\usepackage{slashbox}
\usepackage{algorithm}
\usepackage{algorithmic}
\usepackage{float}
\usepackage{setspace}
\allowdisplaybreaks
\usepackage[a4paper, left=  0.55in, right= 0.55in, top= 0.78 in, bottom= 0.79 in]{geometry}

\makeatletter
\def\widebar{\accentset{{\cc@style\underline{\mskip10mu}}}}
\def\Widebar{\accentset{{\cc@style\underline{\mskip8mu}}}}
\makeatother
\makeatletter
\renewcommand{\maketag@@@}[1]{\hbox{\m@th\normalsize\normalfont#1}}%
\makeatother

\theoremstyle{plain}

\theoremstyle{definition}
\theoremstyle{definition}

\begin{document}
\vspace{-0.1cm}
\title{\LARGE{Rotatable Antenna-Enabled Spectrum Sharing in Cognitive Radio Systems}}
\vspace{-1.0cm}
\author{{\fontsize{10pt}{\baselineskip}\selectfont{
Yanhua Tan, Beixiong Zheng, {\em Senior Member, IEEE}, Yi Fang, {\em Senior Member, IEEE}, Derrick Wing Kwan Ng, {\em Fellow, IEEE}, Jie Xu, {\em Fellow, IEEE}, and Rui Zhang, {\em Fellow, IEEE}}
\vspace{-1.0cm}}
\thanks{Yanhua Tan and Beixiong Zheng are with the School of Microelectronics, South China University of Technology, Guangzhou 511442, China (e-mails: tanyanhua06@163.com; bxzheng@scut.edu.cn).}
\thanks{Yi~Fang is with the School of Information Engineering, Guangdong University of Technology, Guangzhou 510006, China (e-mail: fangyi@gdut.edu.cn).}
\thanks{Derrick Wing Kwan Ng is with the School of Electrical Engineering and Telecommunications, University of New South Wales, Sydney, NSW 2052, Australia (e-mail: w.k.ng@unsw.edu.au).}
\thanks{Jie Xu is with the School of Science and Engineering (SSE), Shenzhen Future Network of Intelligence Institute (FNiiShenzhen), and Guangdong Provincial Key Laboratory of Future Networks of Intelligence, The Chinese University of Hong Kong (Shenzhen), Guangdong 518172, China (E-mail: xujie@cuhk.edu.cn).}
\thanks{Rui Zhang is with the Department of Electrical and Computer Engineering, National University of Singapore, Singapore 117583 (e-mail: elezhang@nus.edu.sg).}

}
\maketitle

\begin{abstract}
Rotatable antenna (RA) technology has recently drawn significant  attention in wireless systems owing to its unique ability to exploit additional spatial degrees-of-freedom (DoFs) by dynamically adjusting the three-dimensional (3D) boresight direction of each antenna. In this letter, we propose a new RA-assisted cognitive radio (CR) system designed to achieve efficient spectrum sharing while mitigating interference  between primary and secondary communication links. Specifically, we formulate an optimization problem for the joint design of the transmit beamforming and the boresight directions of RAs at the secondary transmitter (ST), aimed at maximizing the received signal-to-interference-plus-noise ratio (SINR) at the secondary receiver (SR), while satisfying both interference constraint at the primary receiver (PR) and the maximum transmit power constraint at the ST. Although the formulated problem is challenging to solve due to its non-convexity and coupled variables, we develop an efficient algorithm by leveraging alternating optimization (AO) and successive convex approximation (SCA) techniques to acquire high-quality solutions. Numerical results demonstrate that the proposed RA-assisted system substantially outperforms conventional benchmark schemes in spectrum-sharing CR systems, validating RA's capability to simultaneously enhance the communication quality at the SR and mitigate interference at the PR.
\end{abstract}

\section{Introduction}\label{sect:1 Introduction}
With the rapid proliferation of wireless devices and services, spectrum scarcity has become a critical issue in modern wireless communications. To address this pressing challenge, cognitive radio (CR) has emerged as  a promising paradigm, advocating secondary users (SUs) to access spectrum originally allocated to primary users (PUs) while maintaining the quality of service (QoS) of the PUs~\cite{zhang2010dynamic}. Among the various available spectrum access models in CR systems, spectrum sharing offers a distinctive advantage by allowing SUs to coexist with PUs utilizing the same frequency band while adhering to a prescribed interference constraint, thereby improving spectrum utilization~\cite{10231370}. To facilitate efficient spectrum sharing, numerous advanced techniques have been explored, such as beamforming~\cite{wei2024joint,dong2025joint} and power control~\cite{9031419}. Despite their effectiveness, these techniques primarily rely on conventional fixed-antenna architectures, where antenna positions and orientations remain static post-deployment. This rigidity limits the full exploitation of spatial degrees of freedom (DoFs), which in turn diminishes the system's adaptability to diverse and dynamic environments. Thus, developing more flexible antenna architectures is crucial for harnessing additional spatial DoFs, thereby facilitating more effective spectrum sharing in CR networks.

Recently, rotatable antenna (RA) has attracted great attention as an effective technology for enhancing wireless communication and sensing performance by dynamically adjusting the boresight direction of each antenna to exploit additional spatial DoFs~\cite{wu2024modeling,zheng2025rotatableModeling,zhengtian2025rotatable}. Unlike conventional fixed-antenna arrays, RA arrays provide independent three-dimensional (3D) boresight control for each antenna element, facilitating versatile beam steering and spatial reconfigurability. In particular, RA can be considered as a special case of the six-dimensional movable antenna (6DMA)~\cite{shao20246d,shao2025tutorial} in terms of movement model, retaining the rotational capability only while enabling a more compact design without the need for changing the antenna position in 6DMA. Owing to these advantages, initial studies have been devoted to establishing system and channel models for RA-enabled systems~\cite{wu2024modeling,zheng2025rotatableModeling}, as well as exploring their potential across diverse scenarios~\cite{wu2024modeling,zheng2025rotatableModeling,zhengtian2025rotatable,11098736,zhou2025rotatable,xiong2025intelligent,xiong2025efficient}. For instance, RA has been applied in physical layer security (PLS)~\cite{11098736}, integrated sensing and communication (ISAC)~\cite{zhou2025rotatable}, and more generally integrated sensing, communication, and computation (ISCC)~\cite{xiong2025intelligent}, demonstrating improvements in applications beyond communication.
Notably, through flexible 3D boresight control, RA can selectively enhance or suppress signal power in desired/undesired  directions, thus rendering it well-suited for improving spectrum sharing in CR systems. However, despite these appealing considerations, its potential in such systems has not yet been explored in existing studies.

\begin{figure}[t]
	\vspace{-0.2cm}
	\center
	\includegraphics[width=2.8in,height=1.9in]{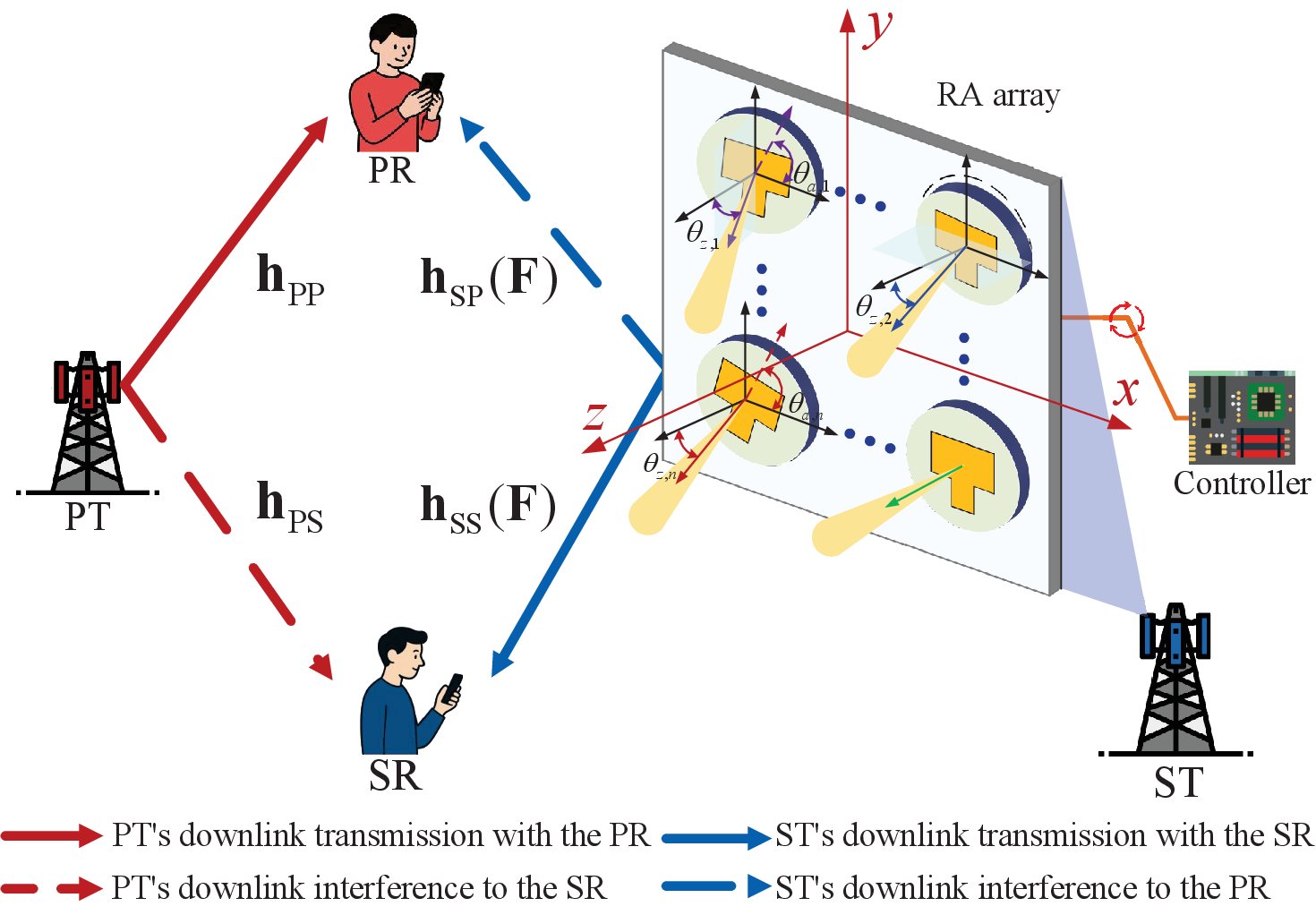}
	\vspace{-0.1cm}
	\caption{An  RA-assisted spectrum-sharing CR system.}\vspace{-0.1cm}
	\label{RA_SS_Eng_v2}  
	\vspace{-0.5cm}
\end{figure}

Motivated by the above insights, this letter investigates the optimization of RAs' boresight directions in a CR system, where an RA array is deployed at the secondary transmitter (ST) to serve a secondary receiver (SR) under a strict interference constraint imposed at the primary receiver (PR), as illustrated in Fig.~\ref{RA_SS_Eng_v2}. To optimize the system performance, we formulate an optimization problem that maximizes the signal-to-interference-plus-noise ratio (SINR) at the SR by jointly optimizing the transmit beamforming and the boresight directions of RAs at the ST. Since the formulated optimization problem is non-convex, we propose an alternating optimization (AO) algorithm that leverages the successive convex approximation (SCA) technique to efficiently achieve a high-quality suboptimal  solution. Numerical results demonstrate that the proposed algorithm achieves significantly better performance than various benchmark schemes.

\section{System Model and Problem Formulation}\label{sect:2 model}
As shown in Fig.~\ref{RA_SS_Eng_v2}, we consider an RA-assisted spectrum-sharing CR system, where a primary transmitter (PT) communicates with a PR, while an ST simultaneously transmits to an SR over the same frequency band.{\footnote{The single-user case is adopted as a baseline model to demonstrate the fundamental performance gains of RAs in spectrum-sharing CR systems. The proposed framework can be readily extended to multiple PRs and/or SRs by incorporating additional interference constraints and multiuser beamforming objectives, which are left for future work.}} In particular, the ST is equipped with a uniform planar array (UPA) consisting of $N$ directional RAs, while the PT is equipped with $M$ traditional fixed antennas that employ optimized transmit beamforming to serve the PR. We assume that both the SR and PR are equipped with a single isotropic antenna. Without loss of generality, we assume that the UPA at the ST is placed on the $x$-$y$ plane of a 3D Cartesian coordinate system and centered at the origin with $N = {N_x}{N_y}$, where $N_x$ and $N_y$ denote the numbers of RAs along the $x$- and $y$-axes, respectively. Let $\mathbf{t}_{\text{ST},n}\in {{\mathbb{R}}^{3\times 1}}, n\in \left\{ 1,2,\cdots ,N \right\}$, ${{\mathbf{t}_{\text{PT},m}}}\in {{\mathbb{R}}^{3\times 1}}, m\in \left\{ 1,2,\cdots ,M \right\}$,
${{\mathbf{t}}_{\text{SR}}}\in {{\mathbb{R}}^{3\times 1}}$, and ${{\mathbf{t}}_{\text{PR}}}\in {{\mathbb{R}}^{3\times 1}}$ denote the positions of the $n$-th RA at the ST, the $m$-th antenna at the PT, the SR, and the PR, respectively.
Moreover, let ${{d}_{q,n}}=\left\| {{\mathbf{t}}_{q}}-\mathbf{t}_{\text{ST},n} \right\|, q\in \left\{\text{SR},\text{PR} \right\}$ be the distance between the $n$-th RA and the SR/PR, where $\parallel\cdot\parallel $ denotes the $\ell_2$-norm.
\subsection{Antenna Boresight Rotation}
The 3D boresight direction of each RA can be mechanically and/or electrically adjusted by a common smart controller~(cf. Fig. 1)~\cite{zheng2025rotatableModeling,zhengtian2025rotatable}.
In particular, the 3D boresight direction of the $n$-th RA  can be characterized by a unit pointing vector, defined as
\begin{equation}\label{pointing vector}
{{\mathbf{f}}_{n}}={{[{{f}_{x,n}},{{f}_{y,n}},{{f}_{z,n}}]}^{T}},
 \end{equation}
where $f_{x,n}$, $f_{y,n}$, and $f_{z,n}$ are the projections of the $n$-th RA's pointing vector on the $x$-, $y$-, and $z$-axes, respectively, and ${{\left( \cdot  \right)}^{T}}$ stands for the transpose operation. The boresight rotation of the $n$-th RA is further described by two angles: the zenith angle $\theta_{z,n}$, defined as the angle between the boresight direction and the $z$-axis, and the azimuth angle $\theta_{a,n}$, defined as the angle between the projection of the boresight direction onto the $x$-$y$ plane and the $x$-axis. Thus, the components of the pointing vector can be expressed as $f_{x,n}=\sin\theta_{z,n}\cos\theta_{a,n}$,
$f_{y,n}=\sin\theta_{z,n}\sin\theta_{a,n},$ and
$f_{z,n}=\cos\theta_{z,n}.$
To incorporate practical rotational limitations~\cite{wu2024modeling,zheng2025rotatableModeling}, the zenith angle of each RA is constrained within a prescribed range, given by $0\le {{\theta }_{z,n}}\le {{\theta }_{{\max}}}$, $\forall n$, where ${{\theta }_{{\max}}}$ denotes the maximum allowable zenith angle for each RA.

\subsection{Channel Model}
In this letter, we consider a generic directional gain pattern for each RA as~\cite{zheng2025rotatableModeling}
\begin{equation}\label{directional gain pattern}
G(\epsilon, \varphi) =
\begin{cases}
G_0 \cos^{2p}(\epsilon), & \epsilon \in [0, \frac{\pi}{2}), \varphi \in [0, 2\pi), \\
0, & \text{otherwise},
\end{cases}
\end{equation}
where $(\epsilon,\varphi)$ denotes a pair of incident angles of the signal with respect to the current boresight direction of the RA, $G_0 = 2(2p +1)$ is the maximum boresight gain satisfying the power conservation law, and $p\ge 0$ denotes the directivity factor. Accordingly, the $n$-th RA's directional gain in the direction of the SR/PR is given by ${{G}_{{q},n}}={{G}_{0}}{{\cos }^{2p}}({{\epsilon }_{{q},n}})$,  where ${{\cos }}({{\epsilon }_{{q},n}})\triangleq {{\mathbf{f}}_{n}}^{T}{{\mathbf{u}}_{{q},n}}$ is the projection between ${{\mathbf{f}}_{n}}^{T}$ and ${{\mathbf{u}}_{{q},n}}$ with ${{\mathbf{u}}_{{q},n}}\triangleq \frac{{{\mathbf{t}}_{{q}}}-\mathbf{t}_{\text{ST},n}}{\left\| {{\mathbf{t}}_{{q}}}-\mathbf{t}_{\text{ST},n} \right\|}$ denoting the unit direction vector from the $n$-th RA to the SR/PR.


For the links from the ST to SR/PR, we consider the near-field line-of-sight (LoS)-dominant channel model as in~\cite{wu2024modeling,zheng2025rotatableModeling}. Specifically, the channel from the $n$-th RA to the SR/PR is given by \begin{equation}\label{channel model}
{{h}_{q,n}}({{\mathbf{f}}_{n}})=\sqrt{g_{q,n}({{\mathbf{f}}_{n}})}{{e}^{-j\frac{2\pi {{d}_{q,n}}}{\lambda }}},\forall q,n,
\end{equation}
where ${{g}_{q,n}}({{\mathbf{f}}_{n}})=\frac{S}{4\pi d_{q,n}^{2}}{{G}_{0}}{{\cos }^{2p}}({{\epsilon }_{q,n}})$ denotes the channel power gain between the $n$-th RA and the SR/PR, $S$ denotes the physical size of each RA element, and $\lambda $ is the signal wavelength. Therefore, the
channel vectors from the ST to the SR and the PR can be expressed as
${{\mathbf{h}}_{\text{SS}}}(\mathbf{F})\triangleq {{\left[ {{h}_{\text{SR},1}}({{\mathbf{f}}_{1}}),{{h}_{\text{SR},2}}(\mathbf{f}_2),...,{{h}_{\text{SR},N}}({{\mathbf{f}}_{N}}) \right]}^{T}}$ and ${{\mathbf{h}}_{\text{SP}}}(\mathbf{F})\triangleq {{\left[ {{h}_{\text{PR},1}}({{\mathbf{f}}_{1}}),{{h}_{\text{PR},2}}(\mathbf{f}_2),...,{{h}_{\text{PR},N}}({{\mathbf{f}}_{N}}) \right]}^{T}}$, respectively, where $\mathbf{F}\triangleq \left[ {{\mathbf{f}}_{1}},{{\mathbf{f}}_{2}},...,{{\mathbf{f}}_{N}} \right]\in {{\mathbb{R}}^{3\times N}}$ is the pointing matrix of all RAs.

\subsection{Problem Formulation}
Let $\mathbf{w}^{H} \in \mathbb{C}^{1 \times N}$ denote the transmit beamforming vector of the ST, subject to the power constraint ${{\left\| \mathbf{w} \right\|}^{2}}\le {{P}_{\max }}$, where $P_{\max}$ is the maximum transmit power of the ST and $(\cdot)^H$ denotes the Hermitian transpose. Similarly, let $\mathbf{v}^{H} \in \mathbb{C}^{1 \times M}$ denote the transmit beamforming vector of the PT, with a constant transmit power  ${{\left\| \mathbf{v} \right\|}^{2}}= {{P}_{0}}$. As a result, the received signals at the
SR and the PR are respectively given by
\begin{align}
 \label{ySR}&{{y}_{\text{SR}}}=\underbrace{{{\mathbf{w}}^{H}}{{\mathbf{h}}_{\text{SS}}}(\mathbf{F}){{x}_{s}}}_{\text{Desired signal}}+\underbrace{{{\mathbf{v}}^{H}}{{\mathbf{h}}_{\text{PS}}}{{x}_{p}}}_{\text{Interference from the PT}}+{{n}_{s}},  \\
 \label{yPR}&{{y}_{\text{PR}}}=\underbrace{{{\mathbf{v}}^{H}}{{\mathbf{h}}_{\text{PP}}}{{x}_{p}}}_{\text{Desired signal}}+\underbrace{{{\mathbf{w}}^{H}}{{\mathbf{h}}_{\text{SP}}}(\mathbf{F}){{x}_{s}}}_{\text{Interference from the ST}}+{{n}_{p}},
\end{align}
where $x_s$ and $x_p$ denote the transmitted signals from the ST and the PT, respectively, such that $\mathbb{E}\left\{ x_{s}^{*}{{x}_{p}} \right\}=0$. Meanwhile, ${\mathbf{h}}_{\text{PS}} \in \mathbb{C}^{M \times 1}$ and ${\mathbf{h}}_{\text{PP}}\in \mathbb{C}^{M \times 1}$ represent the channel vectors from the PT to the SR and to the PR, respectively, while $n_s$ and $n_p$ denote the additive white Gaussian noise (AWGN) at the SR and the PR with zero mean and variances $\sigma_s^2$ and $\sigma_p^2$, respectively. Accordingly, the received SINR at the SR is given by
\begin{equation}\label{SINR}
{{\gamma }_{\text{SR}}}=\frac{{{\left| {{\mathbf{w}}^{H}}{{\mathbf{h}}_{\text{SS}}}({{\mathbf{F}}}) \right|}^{2}}}{{  {{\left| {{\mathbf{v}}^{H}}{{\mathbf{h}}_{\text{PS}}} \right|}^{2}}}+\sigma_s^2}.
\end{equation}

To enable efficient spectrum sharing and guarantee the communication quality of the PR, we consider the interference temperature (IT) constraint, which is commonly adopted in CR systems~\cite{zhang2010dynamic}. Specifically, to mitigate the harmful co-channel interference from the ST to the PR, the received interference power at the PR, given by ${{\left| {{\mathbf{w}}^{H}}{{\mathbf{h}}_{\text{SP}}}(\mathbf{F}) \right|}^{2}}$, must not exceed a prescribed threshold $\Gamma $, i.e.,
 \begin{equation}\label{threshold}
{{\left| {{\mathbf{w}}^{H}}{{\mathbf{h}}_{\text{SP}}}(\mathbf{F}) \right|}^{2}}\le \Gamma.
\end{equation}

In this letter, we jointly optimize the transmit beamforming vector $\mathbf{w}$ and the RA's pointing matrix $\mathbf{F}$ at the ST to maximize the received SINR at the SR, i.e., $\gamma_{\text{SR}}$ in~\eqref{SINR}, while satisfying the IT constraint in~\eqref{threshold}.  Accordingly, the design is formulated as

\vspace{-0.5cm}
\begin{align}
\setcounter{equation}{8}
\label{8a}		(\text{P1}) \quad \max_{\boldsymbol{\mathbf{w},~\mathbf{F}}}&\quad {{\gamma }_{\text{SR}}} \tag{8a}\\  \nonumber\\[-0.65cm]	
\label{8b}		\text{s.t.}  &\quad {{\left\| \mathbf{w} \right\|}^{2}}\le {{P}_{\max }},  \tag{8b}\\
\label{8c}    &\quad  {{\left| {{\mathbf{w}}^{H}}{{\mathbf{h}}_{\text{SP}}}(\mathbf{F}) \right|}^{2}}\le \Gamma,   \tag{8c}\\
\label{8d}    &\quad \text{0}\le \arccos (\mathbf{f}_{n}^{T}{{\mathbf{e}}_{3}})\le {{\theta }_{\max }},\forall n,   \tag{8d}\\
\label{8e}  &\quad  {{\left\| \mathbf{f}_n \right\|}}=1,\forall n,  \tag{8e}
\end{align}
where constraint (8d) limits the rotation of the RAs' boresight directions to a prescribed range with ${{\mathbf{e}}_{3}}\triangleq {{\left[ 0,0,1 \right]}^{T}}$ denoting the unit vector along the $z$-axis, and constraint (8e) ensures that $\mathbf{f}_n$ is a unit vector. To explore the performance limit of the proposed RA-assisted spectrum-sharing CR system, we assume that all necessary channel state information (CSI) is available by applying existing channel estimation techniques dedicated to RAs, e.g.,~\cite{xiong2025efficient}. However, (P1) is difficult to solve optimally due to the non-concave objective function as well as the intricate coupling between the transmit beamforming $\mathbf{w}$ and the RA's pointing matrix $\mathbf{F}$. To efficiently tackle the problem, we propose an AO algorithm that iteratively optimizes the transmit beamforming vector and the pointing matrix.
\vspace{-0.1cm}
\section{Proposed Algorithm for Problem (P1)}
In this section, we develop an AO-based iterative algorithm that alternately solves two subproblems derived from (P1) until convergence is reached.
\subsection{Transmit Beamforming Optimization}
For a given RA's pointing matrix $\mathbf{F}$, the channels from the ST to the SR/PR become fixed. Accordingly, problem (P1) is simplified into
\begin{align}
\setcounter{equation}{9}
(\text{P2}) \quad \max_{\boldsymbol{\mathbf{w}}}&\quad {{\left| {{\mathbf{w}}^{H}}{{\mathbf{h}}_{\text{SS}}}(\mathbf{F}) \right|}^{2}} \tag{9a}\\ \nonumber\\[-0.65cm]	
\text{s.t.} &\quad {{\left\| \mathbf{w} \right\|}^{2}}\le {{P}_{\max }},\tag{9b}\\
&\quad {{\left| {{\mathbf{w}}^{H}}{{\mathbf{h}}_{\text{SP}}}(\mathbf{F}) \right|}^{2}}\le \Gamma ,\tag{9c}
\end{align}
where the noise power ${\sigma_s^2}$ and the interference power ${{\left| {{\mathbf{v}}^{H}}{{\mathbf{h}}_{\text{PS}}} \right|}^{2}}$ in the objective function~\eqref{8a} are omitted. According to~\cite{zhang2008exploiting}, under the single-antenna SR and PR setting, the optimal beamforming vector $\mathbf{w}$ in a multiple-input single-output (MISO) system admits a closed-form solution based on a geometric approach. The optimal beamforming vector is thus given by

\vspace{-0.3cm}
{\footnotesize
\begin{align}\label{w_closen}
{{\mathbf{w}}}=\left\{ \begin{array}{*{35}{l}}
   \sqrt{{{P}_{\max }}}\frac{{{\mathbf{h}}_{\text{SS}}}}{\left\| {{\mathbf{h}}_{\text{SS}}} \right\|}, & \text{ }\frac{{{\left| \mathbf{h}_{\text{SP}}^{H}{{\mathbf{h}}_{\text{SS}}} \right|}^{2}}}{{{\left\| {{\mathbf{h}}_{\text{SS}}} \right\|}^{2}}}\le \frac{\Gamma }{{{P}_{\max }}},  \\
   \sqrt{{{P}_{\max }}}\left( \rho {{{\mathbf{\tilde{h}}}}_{\text{SP}}}+\sqrt{1-{{\rho }^{2}}}\mathbf{\tilde{h}}_{\text{SS}}^{\bot } \right), & \text{ otherwise},  \\
\end{array} \right.
\end{align}}\normalsize where $\mathbf{\tilde{h}}_{\text{SS}}^{\bot }\triangleq \frac{{{\mathbf{h}}_{\text{SS}}}-(\mathbf{\hat{h}}_{\text{SP}}^{H}{{\mathbf{h}}_{\text{SS}}}){{{\mathbf{\hat{h}}}}_{\text{SP}}}}{\left\| {{\mathbf{h}}_{\text{SS}}}-(\mathbf{\hat{h}}_{\text{SP}}^{H}{{\mathbf{h}}_{\text{SS}}}){{{\mathbf{\hat{h}}}}_{\text{SP}}} \right\|}$ denotes the normalized projection of $\mathbf{h}_{\text{SS}}$ onto the nullspace of ${{{\mathbf{\hat{h}}}}_{\text{SP}}}\triangleq \frac{{{\mathbf{h}}_{\text{SP}}}}{\left\| {{\mathbf{h}}_{\text{SP}}} \right\|}$, $\rho \triangleq \sqrt{\frac{\Gamma }{{{P}_{\max }}{{\left\| {{\mathbf{h}}_{\text{SP}}} \right\|}^{2}}}}$
 is the proportion of power allocated in the direction of ${{{\mathbf{\tilde{h}}}}_{\text{SP}}}\triangleq{{e}^{-j\angle (\mathbf{h}_{\text{SS}}^{H}{{{\mathbf{\hat{h}}}}_{\text{SP}}})}}{{{\mathbf{\hat{h}}}}_{\text{SP}}}$, and $\angle (\cdot )$ denotes the phase of a complex scalar.
\subsection{Pointing Matrix Optimization}
In this subsection, we aim to optimize the pointing matrix $\mathbf{F}$ with a given beamforming vector $\mathbf{w}$. For simplicity, let

\vspace{-0.3cm}
{\footnotesize
\begin{equation}\label{J(x)}
J(\mathbf{\overline{f}}) \triangleq  {{\left| {{\mathbf{w}}^{H}}{{\mathbf{h}}_{\text{SS}}}({{\mathbf{\overline{f}}}}) \right|}^{2}}={{\left| \sum\limits_{n=1}^{N}{w_{n}^{*}{{a}_{\text{SR,}n}}{({{{\beta }_{n}})}^{p}}} \right|}^{2}},
\end{equation}}
\vspace{-0.3cm}
{\footnotesize
\begin{equation}\label{U(x)}
U(\mathbf{\overline{f}}) \triangleq {{\left| {{\mathbf{w}}^{H}}{{\mathbf{h}}_{\text{SP}}}({{\mathbf{\overline{f}}}}) \right|}^{2}}={{\left| \sum\limits_{n=1}^{N}{w_{n}^{*}{{a}_{\text{PR,}n}}{{({{\alpha }_{n}})}^{p}}} \right|}^{2}},
\end{equation}}\normalsize
where $\mathbf{\overline{f}}\triangleq \text{vec}(\mathbf{F})\triangleq{{\left[ \mathbf{f}_{1}^{T},\mathbf{f}_{2}^{T},\cdots ,\mathbf{f}_{N}^{T} \right]}^{T}}\in {{\mathbb{R}}^{3N\times 1}}$, $\text{vec}(\cdot)$ represents the vectorization, ${{\beta }_{n}}\triangleq \mathbf{f}_{n}^{T}{{\mathbf{u}}_{\text{SR},n}}$, ${{a}_{\text{SR,}n}}\triangleq\sqrt{\frac{S{{G}_{0}}}{4\pi d_{\text{SR,}n}^{2}}}$, ${{\alpha }_{n}}\triangleq\mathbf{f}_{n}^{T}{{\mathbf{u}}_{\text{PR},n}}$, and ${{a}_{\text{PR,}n}}\triangleq\sqrt{\frac{S{{G}_{0}}}{4\pi d_{\text{PR,}n}^{2}}}$.
To ensure differentiability within the feasible domain, we introduce a small regularization constant $\kappa > 0$~\cite{bertsekas1997nonlinear}. Accordingly, we redefine ${{{\tilde{\alpha }}}_{n}}={{\alpha }_{n}}+\kappa \ge \kappa >0$ and ${{{\tilde{\beta }}}_{n}}={{\beta }_{n}}+\kappa \ge \kappa >0$.{\footnote{ Note that the differentiability of functions $J(\mathbf{\overline{f}})$ and $U(\mathbf{\overline{f}})$ is essential, as the subsequent theoretical analysis relies on well-defined gradients and Hessians.}
Based on~\eqref{J(x)} and~\eqref{U(x)}, the gradients of $J(\mathbf{\overline{f}})$ and $U(\mathbf{\overline{f}})$ with respect to  $\mathbf{f}_n$ are expressed respectively as
\begin{equation}\label{J(x)gradient}
{{\nabla }_{{{\mathbf{f}}_{n}}}}J(\mathbf{\overline{f}})=2\operatorname{Re}\{(A{{(\mathbf{\overline{f}})})^{*}}(p\tilde{\beta} _{n}^{p-1}{{e}_{n}}{{\mathbf{u}}_{\text{SR},n}})\}\in {{\mathbb{R}}^{3\times 1}},
\end{equation}
\begin{equation}\label{U(x)gradient}
{{\nabla }_{{{\mathbf{f}}_{n}}}}U(\mathbf{\overline{f}})=2\operatorname{Re}\{(B{{(\mathbf{\overline{f}}))}^{*}}(p\tilde{\alpha} _{n}^{p-1}{{c}_{n}}{{\mathbf{u}}_{\text{PR},n}})\}\in {{\mathbb{R}}^{3\times 1}},
\end{equation}
where $A(\mathbf{\overline{f}})=\sum\nolimits_{n=1}^{N}{{{e}_{n}}\tilde\beta _{n}^{p}}$, ${{e}_{n}}=w_{n}^{*}{{a}_{\text{SR,}n}}$, $B(\mathbf{\overline{f}})=\sum\nolimits_{n=1}^{N}{{{c}_{n}}{\tilde\alpha} _{n}^{p}}$, ${{c}_{n}}=w_{n}^{*}{{a}_{\text{PR,}n}}$, and {$\rm Re(\cdot)$ is the real part of a quantity.
Since $\mathbf{\overline{f}}$ does not affect the denominator of~\eqref{SINR}, we can omit the constant terms (i.e., ${\sigma_s^2}$ and ${{\left| {{\mathbf{v}}^{H}}{{\mathbf{h}}_{\text{PS}}} \right|}^{2}}$) when maximizing ${{\gamma }_{\text{SR}}}$ with respect to $\mathbf{\overline{f}}$.
Accordingly, (P1) can be simplified as
\begin{align}
\setcounter{equation}{15}
\label{16a}		(\text{P3}) \quad \max_{\boldsymbol{\mathbf{\overline{f}}}}&\quad J(\mathbf{\overline{f}}) \tag{15a}\\ \nonumber\\[-0.65cm]	
\label{16b}		\text{s.t.}  &\quad  U(\mathbf{\overline{f}})\le \Gamma, \tag{15b} \\
\label{16c}	                 &\quad  \cos ({{\theta }_{\max }})\le \mathbf{f}_{n}^{T}{{\mathbf{e}}_{3}}\le 1,\forall n \tag{15c} \\
&\quad {\text{(8e),}}\nonumber
\end{align}
where constraint~\eqref{16c} is equivalent to~\eqref{8d}, restricting each RA's zenith angle within $\left[ 0,{{\theta }_{\max }} \right]$.

However, this subproblem remains challenging to solve due to the non-concavity of the objective function and the non-convexity of constraints (8e) and~\eqref{16b} with respect to the RA's pointing matrix $\mathbf{F}$. To tackle this problem, we adopt the SCA technique to approximate (P3) as a convex problem and obtain a locally optimal solution. Specifically, by
exploiting the first-order Taylor expansion at ${{\mathbf{\overline{f}}}^{(i)}}$, a concave surrogate objective function $J(\mathbf{\overline{f}})$ can be established as
\begin{equation}\label{17}
\tilde{J}(\mathbf{\overline{f}})= J({{\mathbf{\overline{f}}}^{(i)}})+(\nabla J{{({{\mathbf{\overline{f}}}^{(i)}})})^{T}}(\mathbf{\overline{f}}-{{\mathbf{\overline{f}}}^{(i)}}),
\end{equation}
where ${{\nabla }}J(\mathbf{\overline{f}})\triangleq {{\left[ {{({{\nabla }_{{{\mathbf{f}}_{1}}}}J(\mathbf{\overline{f}}))}^{T}},{{({{\nabla }_{{{\mathbf{f}}_{2}}}}J(\mathbf{\overline{f}}))}^{T}},\cdots,{{({{\nabla }_{{{\mathbf{f}}_{n}}}}J(\mathbf{\overline{f}}))}^{T}} \right]}^{T}} \\  \in {{\mathbb{R}}^{3N\times 1}}$ and  ${{\mathbf{\overline{f}}}^{(i)}}$ is the solution of ${{\mathbf{\overline{f}}}}$ at the $i$-th iteration.

Furthermore, to tackle the non-convexity of constraint~\eqref{16b}, we construct a quadratic convex upper bound for $U(\mathbf{\overline{f}})$. A key condition for this construction is that the gradient of $U(\mathbf{\overline{f}})$ is Lipschitz continuous~\cite{bertsekas1997nonlinear} with constant $L_g>0$, as proved in the appendix. Based on this condition and the descent lemma~\cite{bertsekas1997nonlinear}, for any $L\ge L_g$, we have
\begin{equation}\label{18}
U(\mathbf{\overline{f}})\le \tilde{U}(\mathbf{\overline{f}}),
\end{equation}
where {\small$\tilde{U}(\mathbf{\overline{f}})= U({{\mathbf{\overline{f}}}^{(i)}})+\nabla U{{({{\mathbf{\overline{f}}}^{(i)}})}^{T}}(\mathbf{\overline{f}}-{{\mathbf{\overline{f}}}^{(i)}})+\frac{{{L}}}{2}\left\| \mathbf{\overline{f}}-{{\mathbf{\overline{f}}}^{(i)}} \right\|^{2}$} is the quadratic convex upper bound for ${U}(\mathbf{\overline{f}})$. In general, we set $L = L_g$ to obtain a tight upper bound. According to the transitivity of inequalities, if the inequality  $\tilde{U}(\mathbf{\overline{f}})\le \Gamma $ holds, then constraint~\eqref{16b} is guaranteed to be satisfied.
Consequently, a convex subset of~\eqref{16b} is obtained by
 \begin{equation}\label{19}
\tilde{U}(\mathbf{\overline{f}}) \le \Gamma.
\end{equation}

As for the unit-norm constraint for $\mathbf{f}_n$ in~\eqref{8e}, the equality constraint can be relaxed to the following convex set~\cite{bertsekas1997nonlinear}
\begin{equation}\label{20}
{{\left\| \mathbf{f}_n \right\|}} \le 1,\forall n,
\end{equation}
\begin{equation}\label{21}{{ (\mathbf{f}^{(i)}_n)^T \mathbf{f}_n}} \ge 1-\delta,\forall n,
\end{equation}where $\mathbf{f}^{(i)}_n$ denotes the solution of $\mathbf{f}_n$ at the $i$-th iteration and $\delta>0$ is a sufficiently small constant. Following the analysis, problem (P3) is transformed in the $(i{+}1)$-th iteration as follows.
\begin{align*}
\setcounter{equation}{21}
\label{22a}		(\text{P4}) \quad \max_{\boldsymbol{\mathbf{\overline{f}}}}&\quad \tilde{J}(\mathbf{\overline{f}}) \tag{21}\\ \nonumber\\[-0.65cm]	
\label{22b}		\text{s.t.}  &\quad {\text{(15c), (18), (19), (20).}}\nonumber
\end{align*}
Problem (P4) is a convex optimization problem, which can be efficiently solved exploiting the CVX solver.
\subsection{Overall Algorithm}
\begin{algorithm}	
    		\caption{Proposed AO-based Algorithm for (P1).}
            \label{alg1}
    		\begin{algorithmic}[t]
    			\STATE \textbf{Initialize:} Pointing matrix $\mathbf{F}^{(0)} = \begin{bmatrix}
    				\mathbf{e}_3, \cdots, \mathbf{e}_3
    			\end{bmatrix}_{3 \times N}$, threshold $\varepsilon > 0$, and $i = 0$.
    			\REPEAT    			
    			\STATE  Compute $\mathbf{w}^{(i+1)}$ by the closed-form solution in~\eqref{w_closen} with  given $\mathbf{F}^{(i)}$.
    			\STATE Obtain $\mathbf{F}^{(i+1)}$ by solving problem (P4) with  given $\mathbf{w}^{(i+1)}$.
    			\STATE Update $i = i + 1$.
    			\UNTIL{$\left|\frac{{\gamma }_{\text{SR}}^{(i+1)} - {\gamma }_{\text{SR}}^{(i)}}{{\gamma }_{\text{SR}}^{(i)}}\right| \leq \varepsilon$.}
    			\STATE \textbf{Output:} $\mathbf{w} = \mathbf{w}^{(i)}$ and $\mathbf{F} = \mathbf{F}^{(i)}$.
    		\end{algorithmic}
    	\end{algorithm}
To summarize, the overall AO algorithm to solve (P1) is presented in Algorithm 1. In each iteration, the transmit beamforming vector is acquired from the closed-form solution in~\eqref{w_closen}, while the RA's pointing matrix is updated by solving a convex optimization problem. Since this process consistently produces a non-decreasing objective value of (P1) and the objective function is upper bounded, the convergence of Algorithm 1 is guaranteed. In terms of computational complexity, the cost of updating the transmit beamforming vector is negligible, while solving the problem (P4) incurs a complexity of $\mathcal{O}(N^{3.5} \ln(1/\varepsilon))$ per iteration. Hence, the overall computational complexity of Algorithm 1 is $\mathcal{O}(T {N}^{3.5}\ln(1/\varepsilon))$, where $T$ and $\varepsilon$ represent the total number of iterations in the AO algorithm and the accuracy threshold for convergence, respectively.

\section{Numerical Results and Discussions}
In this section, we present simulation results to evaluate the performance of the proposed RA-assisted spectrum-sharing CR system with the proposed AO algorithm. We assume that both the ST and PT are equipped with $N = M = 4$ antennas. The antenna separation is $\Delta =\frac{\lambda }{2}$ with $\lambda=0.125$ meter (m). Moreover, the SR is located at $\big[50\cos\left(\phi\right), 0, 50\sin\left(\phi\right)\big]~\text m$, while the PR and PT are positioned at $[-30, 0, 30]~\text m$ and $[-55, 0, 0]~\text m$, respectively, where $\phi$ denotes the angle between the projection of the SR position onto the $x$-$z$ plane and the $x$-axis. Unless otherwise specified,  we set the angle $\phi=\frac{\pi }{3}$, the ST's maximum transmit power $P_{\max}= 23$ dBm, the PT's transmit power $P_{0} =23$ dBm, the noise power ${{\sigma^{2}_s}}= -80$ dBm, the threshold $\Gamma = -80$ dBm, the directivity factor $p = 4$, and the maximum zenith angle $\theta_{\max} = \frac{\pi}{3}$~\cite{wei2024joint,zheng2025rotatableModeling}.

For a more comprehensive performance evaluation of the proposed RA-assisted spectrum-sharing CR system, referred to as the {\bf{RA-assisted scheme}}, we consider the following three benchmark schemes, all of which employ the optimal beamforming described in Section~III-A, given a typical scenario with fixed PR and SR positions: 1) {\bf{Fixed-antenna scheme}}: The orientations of all RAs are fixed at their reference orientations, i.e., ${{\mathbf{f}}_{n}}={{\mathbf{e}}_{3}},\forall n$. 2) {\bf{Random-orientation scheme}}: The orientation of each RA is randomly generated within the rotational ranges specified in~\eqref{8d} and the simulation results are obtained by averaging over 100 independent random boresight realizations. 3) {\bf{Isotropic-antenna scheme}}: The directional gain is set to $G_0 = 1$ with $p = 0$ in~\eqref{directional gain pattern}.

\begin{figure}[!tbp]
	\center
\vspace{-12mm}
	\includegraphics[width=2.25in,height=1.8in]{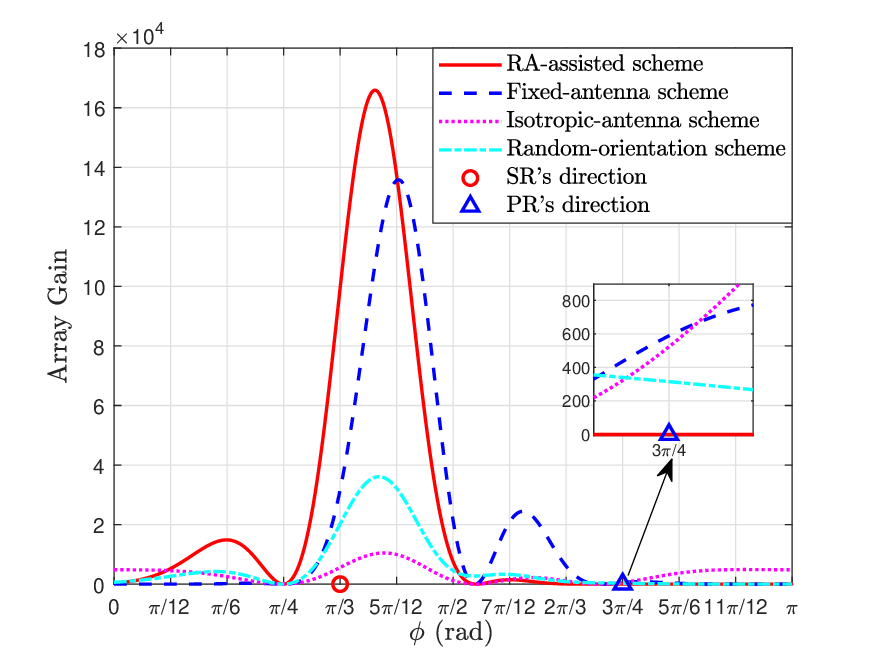}
	\vspace{-0.4cm}
{\caption{Comparison of array gain patterns for different schemes.}
	\label{fig.4}} 
	\vspace{-4mm}
\end{figure}

Fig.~\ref{fig.4} shows the array gain patterns of different schemes versus the angle $\phi \in \left[ 0,\pi  \right]$. The results indicate that all schemes exhibit substantial suppression of array gain toward the PR, validating the effectiveness of the transmit beamforming design. Furthermore, the proposed RA-assisted scheme achieves superior interference suppression at the PR and enhanced array gain at the SR, owing to the additional spatial DoFs enabled by the RA array. In contrast, the fixed-antenna scheme focuses radiation power toward a fixed direction and cannot adapt the boresight direction to serve the SR effectively, rendering it more difficult to satisfy the interference constraint. Similarly, the random-orientation scheme fails to strategically align the antenna's boresight to the desired direction, yielding inferior array gain in the SR direction compared to the proposed RA-assisted scheme. These results validate the effectiveness of the proposed RA-assisted scheme in flexibly reconfiguring the array directional gain pattern to enhance communication quality and mitigate interference in CR systems.

\begin{figure}[!tbp]
	\center
\vspace{-0mm}
	\includegraphics[width=2.25in,height=1.8in]{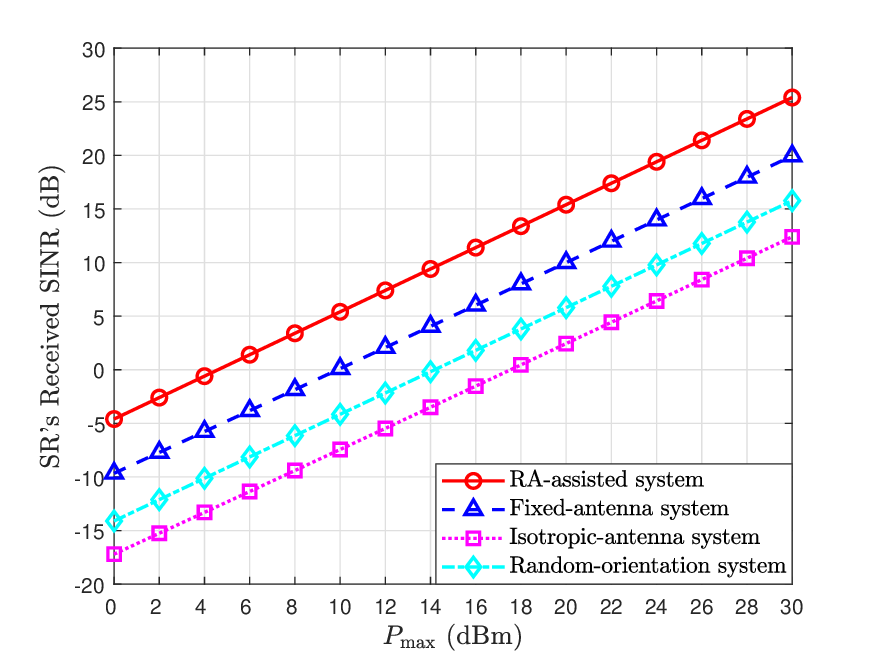}
	\vspace{-0.4cm}
{\caption{The received SINR at the SR versus the maximum transmit power at the ST.}
	\label{fig.2}} 
	\vspace{-6mm}
\end{figure}

Fig.~\ref{fig.2} illustrates the received SINR at the SR versus the maximum transmit power $P_{\max}$ for all considered schemes. As expected, the SINR performance improves with increasing $P_{\max}$ across all schemes. Notably, the proposed RA-assisted scheme consistently delivers superior SINR performance compared to other benchmark schemes. This improvement is primarily attributed to the ability of RAs to adaptively rotate their boresight directions, thereby focusing radiation power toward the SR while effectively mitigating interference toward the PR. In contrast, the SINR performance of the fixed-antenna scheme is lower than that of the proposed RA-assisted scheme. This is because the fixed antennas have static boresight directions and cannot adapt their radiation pattern to the channel environment, resulting in partial energy waste and potential interference leakage to the PR. Moreover, due to the lack of boresight direction optimization, the random-orientation scheme is even worse than the fixed-antenna scheme, further underscoring the necessity of effective boresight control.

\begin{figure}[!tbp]
	\center
\vspace{-0mm}
	\includegraphics[width=2.25in,height=1.8in]{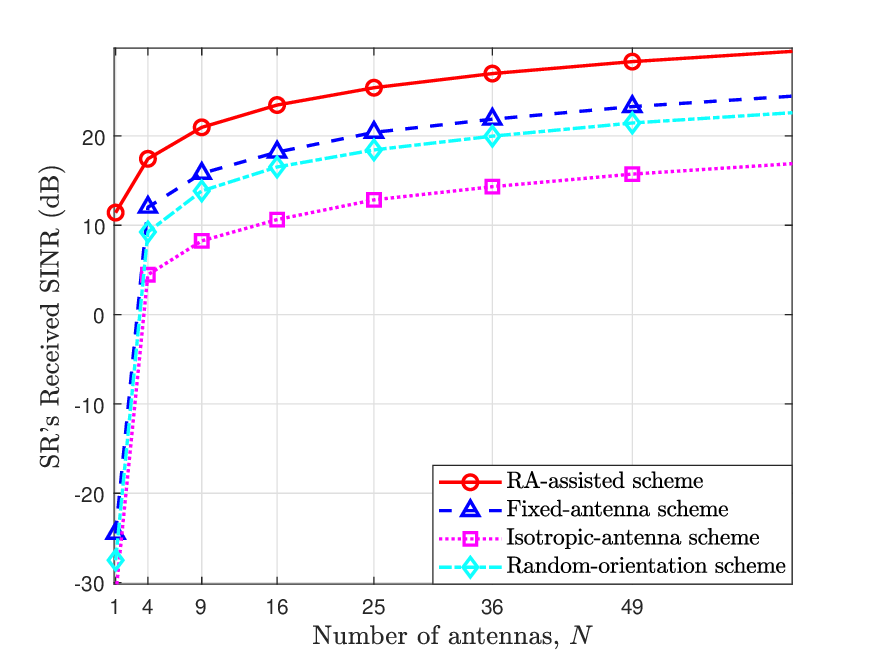}
	\vspace{-0.4cm}
{\caption{The received SINR at the SR versus the number of antennas at the ST.}
	\label{fig.3}} 
	\vspace{-5mm}
\end{figure}

Fig.~\ref{fig.3} depicts the received SINR versus the number of antennas. It is observed that the RA-assisted scheme consistently outperforms all benchmark schemes, and the SINR of each scheme improves as $N$ increases. This trend is attributed to the enhanced array gain brought by a larger antenna array. These results confirm the effectiveness of the proposed RA  boresight optimization algorithm in improving the SINR performance under the strict interference constraint.

\section{Conclusion}\label{sect:conclusion}
In this letter, we proposed a novel RA-assisted spectrum-sharing CR system, where each antenna at the ST can flexibly adapt its boresight direction to rotate the directional gain pattern. By jointly optimizing the transmit beamforming and the boresight directions of RAs, we developed an AO algorithm to enhance the SINR of the SR while harnessing the interference level received at the PR. Simulation results verified that the RA-enabled design outperforms other benchmark schemes in terms of spectrum sharing performance, highlighting its advantages in addressing highly challenging interference-limited scenarios encountered in CR systems.

\section*{Appendix \\Proof of the Lipschitz Continuity of the Gradient $\nabla U(\mathbf{\overline{f}})$
}\label{Appendix}
First, by computing the second-order derivative of the function $U(\mathbf{\overline{f}})$ with respect to $\mathbf{\overline{f}}$, we obtain the Hessian matrix ${{\nabla }^{2}}U(\mathbf{\overline{f}})$, given by

\vspace{-0.3cm}
{\footnotesize
\begin{equation}
{{\nabla }^{2}}U(\mathbf{\overline{f}})=\left[ \begin{matrix}
   {{\mathbf{H}}_{1,1}} & {{\mathbf{H}}_{1,2}} & \cdots  & {{\mathbf{H}}_{1,N}}  \\
   {{\mathbf{H}}_{2,1}} & {{\mathbf{H}}_{2,2}} & \cdots  & {{\mathbf{H}}_{2,N}}  \\
   \vdots  & \vdots  & \ddots  & \vdots   \\
   {{\mathbf{H}}_{N,1}} & {{\mathbf{H}}_{N,2}} & \cdots  & {{\mathbf{H}}_{N,N}}  \\
\end{matrix} \right] \in {{\mathbb{R}}^{3N\times 3N}},
\end{equation}}\normalsize where the Hessian block ${{\mathbf{H}}_{\tilde{n},n}}$ is computed as

\vspace{-0.3cm}
{\footnotesize
\begin{align}\label{eq:H_blocks}
\mathbf{H}_{\tilde{n},n}
=
\begin{cases}
2\operatorname{Re}\Bigl\{
  \bigl(p\,\tilde{\alpha}_{n}^{p-1}c_{n}\mathbf{u}_{{\text {PR}},n}\bigr)
  \bigl(p\,\tilde{\alpha}_{n}^{p-1}c_{n}\mathbf{u}_{{\text {PR}},n}\bigr)^{T}\Bigr.\\
  \Bigl.+B(\mathbf{\overline{f}})^{T}\,
   p(p-1)\,\tilde{\alpha}_{n}^{p-2}\,c_{n}
   \mathbf{u}_{{\text {PR}},n}\mathbf{u}_{{{\text {PR}},n}}^{T}
\Bigr\},
& \tilde{n}=n,\\[0.5em]
2\operatorname{Re}\Bigl\{
  \bigl(p\,\tilde{\alpha}_{n}^{p-1}c_{n}\mathbf{u}_{{{\text {PR}},n}}\bigr)
  \bigl(p\,\tilde{\alpha}_{\tilde{n}}^{p-1}c_{\tilde{n}}\mathbf{u}_{{{\text {PR}},}\tilde{n}}\bigr)^{T}
\Bigr\},
& \tilde{n}\neq n,
\end{cases}
\end{align}}\normalsize with $\tilde {n}\in \left\{ 1,2,\cdots ,N \right\}$. Accordingly, the Hessian block ${{\mathbf{H}}_{\tilde{n},n}}$ satisfies

\vspace{-0.4cm}
{\footnotesize
\begin{align}\label{eq:H_blocks N}
\left\{ \begin{array}{*{35}{l}}
   \left\| {{\mathbf{H}}_{n,n}} \right\|\le 2{{p}^{2}}{{\left| {{c}_{n}} \right|}^{2}}\tilde{\alpha }_{n}^{2(p-1)}+2|p(p-1)|\left| {{c}_{n}} \right|\sum\limits_{i=1}^{N}{\left| {{c}_{i}} \right|}\tilde{\alpha }_{n}^{p-2}, & \tilde{n}=n,  \\
   \left\| {{\mathbf{H}}_{\tilde{n},n}} \right\|\le 2{{p}^{2}}\left| {{c}_{n}} \right|\left| {{c}_{{\tilde{n}}}} \right|\tilde{\alpha }_{n}^{p-1}\tilde{\alpha }_{{\tilde{n}}}^{p-1}, & \tilde{n}\ne n.  \\
\end{array} \right.
\end{align}}\normalsize Since ${{{\tilde{\alpha }}}_{n}}\in \left[ \kappa ,1+\kappa  \right]$, the norm of the Hessian matrix $\nabla^2 U(\mathbf{\overline{f}})$ can be upper bounded as

\vspace{-0.4cm}
{\footnotesize
\begin{align}\label{eq:Lg}
\left\| {{\nabla }^{2}}U(\mathbf{\overline{f}}) \right\| &\le \underset{{\tilde{n}}}{\mathop{\max }}\,\sum\limits_{n=1}^{N}{\left\| {\mathbf{H}_{\tilde{n},n}} \right\|} \nonumber \\
&=2p(p+|p-1|){{C}_{\max }}{{C}_{\text{sum}}}\times
\left\{ \begin{array}{*{35}{l}}
   {{(1+\kappa )}^{2(p-1)}}, & p\ge 1,  \\
   {{\kappa }^{p-2}}, & 0<p<1,  \\
\end{array} \right.\nonumber \\
&\triangleq {{L}_{g}},
\end{align}\normalsize where ${{C}_{\max }}=\underset{n}{\mathop{\max }}\,\left| {{c}_{n}} \right|$,  ${{C}_{\text{sum }}}=\sum\nolimits_{i=1}^{N}{\left| {{c}_{i}} \right|}$, and $L_g>0$.

As a result, for any $\mathbf x, \mathbf y\in {{\mathbb{R}}^{3N\times 1}}$, there exists a point $\mathbf z\in {\mathbb{R}}^{3N\times 1}$  on the line segment connecting $\mathbf x$ and $\mathbf y$ such that

\vspace{-0.3cm}
{\footnotesize
\begin{align}
\|\nabla U(\mathbf{y})-\nabla U(\mathbf{x})\|&=\left\| {{\nabla }^{2}}U(\mathbf{z})(\mathbf{y}-\mathbf{x}) \right\| \nonumber \\
&\le \left\| {{\nabla }^{2}}U(\mathbf{z}) \right\|\|\mathbf{y}-\mathbf{x}\|
\le L_g\|\mathbf{y}-\mathbf{x}\|.
\end{align}\normalsize
Hence, $\nabla U(\mathbf{\overline{f}})$ is Lipschitz continuous with constant $L_g > 0$.

\bibliographystyle{IEEEtran}
\bibliography{IEEEabrv,Reference}

\end{document}